\documentclass[prb,floatfix,showpacs,amssymb,twocolumn]{revtex4}

\usepackage{graphicx}\usepackage{dcolumn}
\usepackage{bm}
\begin{document}

\title{Subgap tunneling via
quantum-interference effect: insulators and charge density waves 
}

\author{S. Duhot}
\affiliation{Institut NEEL, CNRS and Universit\'e Joseph Fourier,
BP 166, F-38042 Grenoble Cedex 9, France}

\author{R. M\'elin}
\affiliation{Institut NEEL, CNRS and Universit\'e Joseph Fourier,
BP 166, F-38042 Grenoble Cedex 9, France}

\begin{abstract}
A quantum interference effect is discussed for
subgap tunneling over a distance comparable to the
coherence length, which is a consequence of
``advanced-advanced'' and
``retarded-retarded'' transmission modes [Altland and Zirnbauer,
Phys. Rev. B {\bf 55}, 1142 (1997)]. Effects typical of disorder
are obtained from the interplay between
multichannel averaging and higher order processes in the tunnel
amplitudes.
Quantum interference effects similar to those occurring in
normal tunnel junctions
explain {\it magnetoresistance
oscillations of  a CDW} pierced by nanoholes 
[Latyshev {\it et al.}, Phys. Rev. Lett. {\bf 78}, 919 (1997)],
having periodicity $h/2e$ as a function of the flux enclosed
in the nanohole.
Subgap tunneling is coupled to the sliding motion
by charge accumulation in the interrupted chains.
The effect is 
within the  same
trend as random matrix theory for normal metal-CDW
hybrids [Visscher {\it et al.},
Phys. Rev. B {\bf 62}, 6873 (2000)]. We suggest that 
the experiment by Latyshev {\it et al.}
probes weak localization-like properties of
evanescent quasiparticles, not an
interference effect related to the 
quantum mechanical ground state.
\end{abstract}

\maketitle

\section{Introduction}
Normal electron tunneling \cite{Caroli} through a barrier is realized 
in two-terminal devices
at the tip of a
scanning tunneling microscope, or by extended interfaces
for planar tunnel junctions. 
Diffusive conductors can be described by arrays of tunnel
junctions \cite{Zaikin,Blanter}
and it is desirable to develop a thorough understanding 
of all aspects of weak localization already at the level 
of a single tunnel junction. Starting from 
a model system of normal electron tunneling through a band insulator
(with hypothesis on the dimensionality and on the 
extrema of the dispersion relation),
we show that weak localization-like subgap tunneling
\cite{M-PRB,DM}
is the clue to an
experiment by Latyshev {\it et al.}\cite{Latyshev}
on $h/2e$ oscillations of the magnetoresistance related to
CDW motion around a
nanohole of size comparable to the CDW coherence length.

CDWs realize a well known phase of quasi-one dimensional (quasi-1D)
conductors \cite{Gorkov}, with density modulations along
the direction of the chains in the ground state,
with a gap and a collective
sliding motion above the depining
threshold. CDWs can be nano-fabricated, as shown by
various experiments in the last decade
\cite{Neill,nanowire,array,AR,Wang,Kasatkin,Artemenko-Andreev}.

\begin{figure*}
\includegraphics [width=.7 \linewidth]{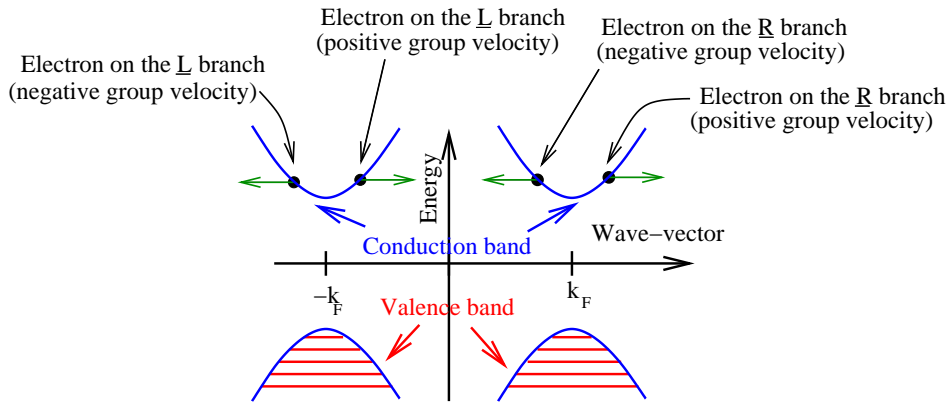}
\caption{(Color online.) Schematic representation of the
BCS-like dispersion relation of the band insulator
with a gap $\Delta$. All states
are populated by genuine electrons and holes, resulting in
a band insulator, not in a superconductor. 
Electrons with wave-vectors $\simeq \pm k_F$ are labeled
by $\underline{R}$ and $\underline{L}$
respectively. A $\underline{R}$ electron
can have a
positive or negative group velocity, as shown on the figure.
\label{fig:disper}
\label{fig:band}
}
\end{figure*}
A film of the CDW compound NbSe$_3$ pierced by the
nanoholes formed by columnar defects was obtained
from ion irradiation 
by Latyshev {\it et al.} \cite{Latyshev}, and the
experimental results were reproduced
after publication.
The diameter $D\simeq 10\,$nm of the nanohole is comparable to
the ballistic CDW coherence length $\xi_0=\hbar v_F/\Delta$,
with $v_F$ the Fermi velocity and $\Delta$ the Peierls gap
of the CDW\cite{Latyshev}.
We reach an agreement with
the following experimental
observations \cite{Latyshev} for the (un)irradiated sample
(not) containing nanoholes: 1. Absence of magnetoresistance oscillations
without nanoholes; 2. Absence of magnetoresistance oscillations
with nanoholes but without sliding motion; 3. $h/2e$ oscillations of the
resistance as a function of magnetic flux
with nanoholes and with sliding motion;
4. Positive magnetoresistance at low fields with nanoholes;
5. Oscillations measured by Latyshev {\it et al.}\cite{Latyshev}
at temperature as high as $\simeq 52$K in the presence of nanoholes.

Previous approaches to related phenomena in CDWs
were based on weak localization in normal metal-CDW hybrids 
in the framework of
random matrix theory \cite{Visscher},
Aharonov-Bohm oscillations \cite{Bogachek},
soliton tunneling \cite{Bogachek,Miller} and
permanent currents \cite{Mon-CDW,Yi}.
The effect that we consider is not
directly related to non linearities of the
CDW phase Hamiltonian
\cite{ps1,ps2,ps3,ps4,ps5,ps6,ps7,ps8,ps9,ps10,ps11,ps12,ps13,Artemenko}.
We reach consistency with
Visscher {\it et al.} \cite{Visscher} finding on the basis of random
matrix theory
an unexpected strong effect of
weak localization for CDWs connected to a disordered normal electrode. 
The effects discussed below rely on a
geometrical parameter being comparable to the coherence length,
namely $D\sim\xi_0$ for a nanohole of diameter $D$
in a CDW film.
By analogy,
transport in superconducting hybrids having a
dimension comparable to the superconducting coherence length
has aroused considerable interest
recently\cite{Lambert,Jedema,Byers,Deutscher,Falci,Samuelson,Prada,Koltai,japs,Feinberg-des,Melin-Feinberg-PRB,Russo,Beckmann,Chandra,Choi,Martin,Peysson,Buttiker,Zaikin-CAR,Golubov,Giazotto,Belzig,Levy}
in connection with the realization of a source of entangled pairs of
electrons.

Non standard localization effects were
already introduced in
superconducting hybrid structures by Altland and Zirnbauer 
for an Andreev quantum dot\cite{Altland}, who state 
at the end of their abstract that in normal metal-superconductor hybrids,
{\it ``every
Cooperon and diffusion mode in the advanced-retarded channel entails a 
corresponding mode in the advanced-advanced (or retarded-retarded)
channel''}, which constitutes the technical basis of part of our discussion
where ``diffusion modes'' become ``transmission modes''.

The structure of the article is as follows.
Preliminaries are presented in Sec.~\ref{sec:prelim}.
Sec.~\ref{sec:PRELIM} is focused 
on tunneling across band insulators.
Sec.~\ref{sec:phys} presents our results on
charge density waves in connection with Sec.~\ref{sec:PRELIM} 
and with the
experiments by Latyshev {\it et al.}\cite{Latyshev}.
Concluding remarks are presented in Sec.~\ref{sec:conclu}.
The discussion in the main body of the article is
complemented by two Appendices.

\section{Preliminaries: Hamiltonians, Green's functions,
weak localization loops}
\label{sec:prelim}
We start with necessary preliminaries on the microscopic description.
Insulators (Sec.~\ref{sec:prelim-ins}) and
charge density waves (Sec.~\ref{sec:prelim-CDW})
are described by the same formalism. Weak localization-like
subgap tunneling is introduced in Sec.~\ref{sec:wl-intro}.

\subsection{Description of a band insulator}
\label{sec:insul-asump}
\label{sec:ins1}
\label{sec:prelim-ins}
We consider for the band insulator the same
dispersion relation as for the BCS model (see Fig.~\ref{fig:disper}),
but assume excitation branches
supporting either electrons or holes, without condensate and
without explicit CDW or
superconducting symmetry breaking.

The advanced Green's function connecting
two sites $\alpha$ and $\beta$ separated by a distance
$R_{\alpha,\beta}$ is given by
\begin{eqnarray}
\label{eq:gA-ins}
g^A_{\alpha,\beta}(\omega)&=&\frac{\pi\rho_N}{k_F |R_{\alpha,\beta}|}
\sin{(k_F |R_{\alpha,\beta}|)}\times\\
&&\frac{-\hbar \omega}{\sqrt{|\Delta|^2-(\hbar\omega)^2}}
\exp{\left(-|R_{\alpha,\beta}|/\xi(\omega)\right)}
\nonumber
,
\end{eqnarray}
where $\rho_N$ is the normal density of states, $k_F$ the Fermi
wave-vector corresponding to the extrema of the dispersion relation,
and $\xi(\omega)=\hbar v_F/\sqrt{|\Delta|^2-(\hbar \omega)^2}$
the coherence length at energy $\hbar \omega$, with
$v_F$ the Fermi velocity.

In addition,
the insulator will be supposed to be quasi-1D
in the forthcoming
Sec.~\ref{sec:q1D}. Then, we separate right
(label $\underline{R}$) from left (label $\underline{L}$)
branches (underlines
are used to avoid confusion with the ``A'' and ``R'' labels
for advanced and retarded Green's functions) having a wave-vector
$\simeq \pm k_F$. An electron on the $\underline{R}$ branch
can have a positive or negative group velocity (see Fig.~\ref{fig:disper}).
The corresponding $2\times2$ matrix Green's function 
between times $t$ and $t'$ and positions $\alpha$ and $\beta$
are given by
\cite{Artemenko-Volkov} 
\begin{eqnarray}
\label{eq:AV}
&&\hat{g}_{\alpha,\beta}^A(t,t') =-i \theta(t-t') \times\\
\nonumber
&&\left[ 
\begin{array}{cc}
\langle \left\{c_{\alpha,\underline{R}}^+(t'),c_{\beta,\underline{R}}(t) 
\right\} \rangle
&
\langle \left\{c_{\alpha,\underline{R}}^+(t'),c_{\beta,\underline{L}}(t) 
\right\} \rangle
\\
\langle \left\{c_{\alpha,\underline{L}}^+(t'),
c_{\beta,\underline{R}}(t) \right\} \rangle
&
\langle \left\{c_{\alpha,\underline{L}}^+(t'),
c_{\beta,\underline{L}}(t) \right\} 
\rangle
\end{array} \right]
\nonumber
.
\end{eqnarray}
After Fourier transform,
Eq.~(\ref{eq:AV}) reduces to
\begin{eqnarray}
\label{eq:g4x4}
&&\hat{g}^A_{\alpha,\beta}(\omega) =
\pi \rho_N \frac{-\hbar\omega}{\sqrt{|\Delta|^2-(\hbar\omega)^2}}
\exp{(-|R_{\alpha,\beta}|/\xi)} \\
&&\times \left[\begin{array}{cc}
\exp{(i k_F R_{\alpha,\beta})} & 0 \\
0 & \exp{(-i k_F R_{\alpha,\beta})} 
\end{array} \right]
\nonumber
\end{eqnarray}
for propagation at energy $\hbar \omega$ between
two points separated by $R_{\alpha,\beta}$ along a given chain.
Eq.~(\ref{eq:g4x4}) in 1D
is compatible with Eq.~(\ref{eq:gA-ins}) in 3D because
the $\sin{(k_F |R_{\alpha,\beta}|)}/k_F |R_{\alpha,\beta}|$
factor in 3D is replaced
by $\cos{(k_F R_{\alpha,\beta})}$ in 1D.

A coupling to the vector potential due to a magnetic field is
obtained by replacing $\hat{g}^A_{\alpha,\beta}(\omega)$ 
in Eq.~(\ref{eq:AV}) by
\begin{equation}
\label{eq:subs}
\hat{g}^A_{\alpha,\beta}(\omega) \exp{\left(i\frac{2\pi}{\phi_0}
\int_\alpha^\beta
{\bf A}\cdot d{\bf r}\right)}
,
\end{equation}
where $\int_\alpha^\beta
{\bf A}\cdot d{\bf r}$ is the circulation of the vector potential on
a path connecting $\alpha$ to $\beta$, and $\phi_0=h/e$ is the flux
quantum.

\subsection{Description of a charge density wave}
\label{sec:prelim-CDW}
\subsubsection{Peierls Hamiltonian}
\label{app:HP}
Charge density waves are described on the basis 
of the electronic part of the
Peierls Hamiltonian of spinless fermions on a 1D chain:
\begin{equation}
\label{eq:H-Peierls}
{\cal H}_P=-\sum_x \left[T+\delta_0 \cos{(2k_F x)}\right]
\left[ c_{x+a_0}^+ c_x + c_x^+ c_{x+a_0}\right]
,
\end{equation}
where the average hopping amplitude $T$ has a modulation $\delta_0$
($a_0$ is
the lattice parameter).
Changing variable from the spatial coordinate $x$ along the
chain to the wave-vector $k$, we find
\begin{eqnarray}
{\cal H}&=&-\sum_k 2 T \cos{(k a_0)} c_k^+ c_k\\
\nonumber
&-&\sum_k \Delta c_{k-2k_F}^+ c_k
-\sum_k \Delta^* c_{k+2k_F}^+ c_k
,
\nonumber
\end{eqnarray}
with $|\Delta|=\delta_0$ the Peierls gap.
\begin{figure}
\includegraphics [width=.85 \linewidth]{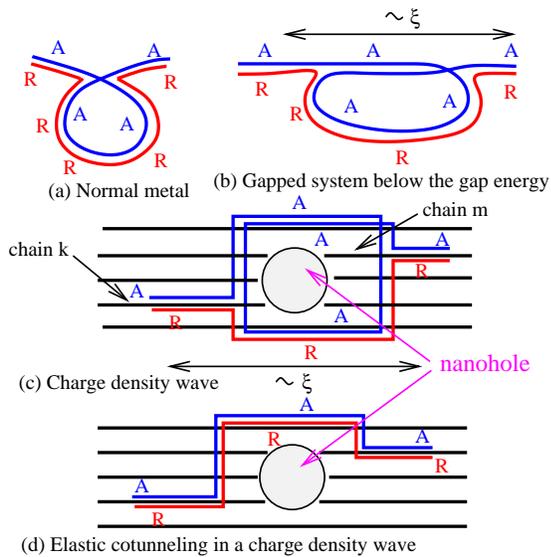}
\caption{(Color online.) Schematic representation in
real space of a
localization loop inserted in diagrams for the
transmission coefficient in a normal metal (a);
and
in a gapped system
below the gap (b).
Similar diagrams are obtained from the tunnel
terms at the interfaces\cite{M-PRB} to higher orders.
(c) shows a weak localization-like loop
for a quasi-1D charge density wave chains pierced by a nanohole
(see Sec.~\ref{sec:phys}). (d) shows a process
now winding around the hole, and therefore not
being modulated by a magnetic flux in the nanohole.
The labels
``A'' and ``R'' stand for
advanced and retarded Green's functions.
\label{fig:loop}
}
\end{figure}

\subsubsection{Ballistic Green's functions}
\label{app:Green}
Ballistic
propagation
within a given CDW chain is described by Eq.~(\ref{eq:AV}).
The CDW Green's functions take the form
\begin{eqnarray}
\label{eq:g-generic}
&&\hat{g}_{\alpha,\beta}^{A}(\omega)= 
\left[\begin{array}{cc} A & B \\
C & D
 \end{array} \right]
,
\end{eqnarray}
with
\begin{eqnarray}
A &=& g_0(\omega) \exp{\left(i k_F (x_\alpha-x_\beta)\right)}\\
B &=& f_0(\omega) \exp{\left(ik_F(x_\alpha+x_\beta)\right)} \\
C &=& f_0(\omega)  \exp{\left(-ik_F(x_\alpha+x_\beta)\right)}\\
D &=& g_0(\omega) \exp{\left(-ik_F(x_\alpha-x_\beta)\right)}
,
\end{eqnarray}
where the points $\alpha$ and $\beta$ are
at coordinates $x_\alpha$ and $x_\beta$ along the chains, and
where
\begin{eqnarray}
\label{eq:g0}
g_0(\omega)&=&{1\over 4T} \left(\frac{-\hbar \omega}
{\sqrt{|\Delta|^2-(\hbar\omega)^2}}+i\right)\\
f_0(\omega)&=&{1\over 4T} \frac{\Delta}
{\sqrt{|\Delta|^2-(\hbar\omega)^2}}
\label{eq:f0}
.
\end{eqnarray}

\subsubsection{Tunneling self-energy}
\label{sec:hopping}
One dimensional chains with an average intra-chain hopping $T$ are
supposed to be connected by a weak interchain hopping $t_\perp$.
Interchain hopping 
corresponds to $c_a^+c_\alpha$, encoding
the destruction of a spinless fermion at site ``$\alpha$'' and
its creation at site $a$ in a neighboring chain, 
as well as to the reversed process.
The fermion creation operator
$c_a^+$ is decomposed
in the right ($\underline{R}$)- and left ($\underline{L}$)-moving
components 
$c_{a,\underline{R}}^+$ and $c_{a,\underline{L}}^+$ 
according to
\begin{equation}
c_a^+= \frac{1}{\sqrt{2}} \left(c_{a,\underline{R}}^+ + 
c_{a,\underline{L}}^+\right)
,
\end{equation}
with
$c_{a,\underline{R}}^+=e^{ik_F x_a} \hat{\chi}_{\underline{R}}^+(x_a)$ and
$c_{a,\underline{L}}^+= e^{-ik_F x_a} \hat{\chi}_{\underline{L}}^+(x_a)$
(site ``$a$'' is at coordinate $x_a$ along the chain).
The fields $\hat{\chi}_{\underline{R}(\underline{L})}^+(x)$ 
are slowly
varying as a function of the coordinate $x$ along the chain.
The identity
\begin{equation}
c_a^+ c_\alpha=\frac{1}{2} \left( c_{a,\underline{L}}^+ 
c_{\alpha,\underline{L}}
+c_{a,\underline{R}}^+ c_{\alpha,\underline{L}} + c_{a,\underline{L}}^+ 
c_{\alpha,\underline{R}}
+c_{a,\underline{R}}^+ c_{\alpha,\underline{R}} \right)
\end{equation}
leads to the self-energy
\begin{equation}
\label{eq:Sigmat}
\hat{\Sigma}_{\alpha\rightarrow a}^{A}=\frac{t_\perp}{2}
\left[\begin{array}{cc}
1 & 1 \\
1 & 1
\end{array} \right]
\end{equation}
for hopping from $\alpha$ to $a$,
where the entries correspond to different
left and right components.
\begin{figure*}
\includegraphics [width=.7 \linewidth]{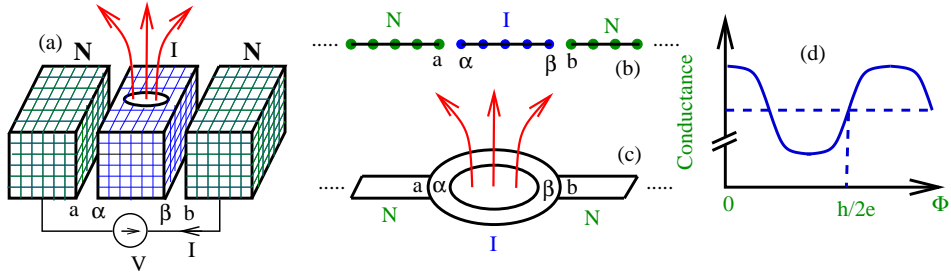}
\caption{(Color online.) Schematic representation of (a) a tight-binding
normal metal-insulator-normal metal (NIN) tunnel junction with a hole
pierced by a magnetic flux $\Phi$ in the insulator (b) the
1D NIN tunnel junction\cite{Caroli}, and (c) a  geometry with
an insulating ring pierced by a flux $\Phi$, and (d)
the resulting $h/2e$ oscillations of the
conductance with a negative magnetoconductance at low field.
The notations $a$, $\alpha$, $\beta$ and $b$ for
labeling the interfaces are shown on the figure.
\label{fig:NIN-ext}
}
\end{figure*}

\subsection{Weak localization-like loops}
\label{sec:wl-intro}
A weak localization loop in a normal metal is
shown on Fig.~\ref{fig:loop}a. The generalization to
subgap transport \cite{DM}
involves a Hikami box\cite{GLK,H} containing an
``advanced-advanced'' or a ``retarded-retarded'' 
transmission mode\cite{Altland}.
Fig.~\ref{fig:loop}b shows a weak localization-like diffuson
self-crossing in a 3D gapped system (such as a
superconductor\cite{DM,Smith})
below
the gap energy, and 
the same type of
diagram from higher order terms in the tunnel amplitudes
in normal metal-insulator-normal metal (NIN)
structures are discussed in the forthcoming 
Sec.~\ref{sec:PRELIM}.
The generalization to quasi-1D CDWs 
(see Sec.~\ref{sec:phys})
corresponds to Fig. \ref{fig:loop}c, and a process
of cotunneling through a CDW is shown on Fig.~\ref{fig:loop}d.
As a direct consequence of Eq.~(\ref{eq:subs}), the transmission
mode associated to Fig.~\ref{fig:loop}b and Fig.~\ref{fig:loop}c
is $h/2e$-periodic
as a function of the flux $\Phi$ enclosed in the loop
(see the discussion in Secs.~\ref{sec:PRELIM} and~\ref{sec:phys}).
The magnetic
field phase factor [see Eq.~(\ref{eq:subs})]
accumulated by the ``advanced'' and ``retarded''
Green's functions cancel with each other in the diagram
on Fig.~\ref{fig:loop}d, which is thus not modulated
by a magnetic field.

\section{Tunneling through a band insulator}
\label{sec:PRELIM}
\label{sec:jonctiontunnel}
We start with a tunnel junction
consisting of a band insulator
inserted in between two normal electrodes.
The assumption on the dispersion relation
(see Fig.~\ref{fig:band} and Sec.~\ref{sec:ins1}) is 
representative of
extrema of the dispersion relation at wave-vectors $p_n$,
with $2\pi/p_n$ much smaller than the thickness of
the insulating layer traversed by tunneling electrons.


\label{sec:tunnel}

\subsection{Ring geometry within microscopic Green's function}
\label{sec:IID}
\label{sec:ring}
\label{sec:IIIA}

\subsubsection{Transport formula}

We start with the Caroli, Combescot, Nozi\`eres 
and Saint-James\cite{Caroli}
expression of the conductance ${\cal G}(\omega)$ of a
1D N$_a$IN$_b$ junction:
\begin{eqnarray}
\label{eq:G1D}
\label{eq:Nozieres}
&&{\cal G}(\omega)=\\
&&4\pi^2 \frac{e^2}{h} 
\rho_{a,a}(\omega) t_{a,\alpha} G_{\alpha,\beta}^A(\omega) t_{\beta,b}
\rho_{b,b}(\omega) t_{b,\beta} G_{\beta,\alpha}^R(\omega) t_{\alpha,a}
\nonumber
,
\end{eqnarray}
where $\hbar \omega$ is equal to the bias voltage energy $e V$; 
$\rho_{a,a}(\omega)$ and $\rho_{b,b}(\omega)$ denote the density of
states in electrodes ``$a$'' and ``$b$'' at energy $\hbar \omega$;
$t_a\equiv
t_{a,\alpha}=t_{\alpha,a}$
and $t_b\equiv t_{b,\beta}=t_{\beta,b}$ are the hopping 
amplitudes for the contact with the points $\alpha$ and
$\beta$ at the extremities of the insulator to their
counterparts $a$ and $b$ in the normal electrodes
(see Figs.~\ref{fig:NIN-ext}a for planar interfaces, and 
Fig.~\ref{fig:NIN-ext}b for the 1D model\cite{Caroli}).
Eq.~(\ref{eq:G1D}) is valid to all orders in the tunnel
amplitudes $t_a$ and $t_b$, from tunnel junctions to
highly transparent interfaces.
The fully dressed advanced and retarded
Green's functions $G_{\alpha,\beta}^A(\omega)$
and $G_{\beta,\alpha}^R(\omega)$ describe propagation from
$\alpha$ to $\beta$ and from
$\beta$ to $\alpha$ respectively while including all possible
excursions in the normal electrodes, as opposed to the notations
$g_{\alpha,\beta}^A(\omega)$
and $g_{\beta,\alpha}^R(\omega)$
restricted to electrodes isolated from each other, with
$t_{a}=t_{b}=0$.
Physically, electron propagation  across the insulator
from electrode N$_a$ at time $\tau_a$ to electrode N$_b$ at time
$\tau_b$ [corresponding to the advanced Green's function
$G_{\alpha,\beta}^A(\omega)$] occurs within the same process as
propagation of an electron backward in time from electrode N$_b$
at time $\tau_b$ to electrode N$_a$ at time $\tau_a$
[corresponding to the retarded Green's function
$G_{\beta,\alpha}^R(\omega)$]. As we show, both electrons
propagating forward and backward in time
can bounce independently at the interfaces during a tunneling process,
corresponding to contributions to the conductance of higher order in 
tunnel amplitudes. This amounts to evaluating a
transition probability as the square of a transition amplitude,
which contains interference terms corresponding among others to
weak localization-like subgap tunneling.

\subsubsection{Expansion of the transport formula}
\label{sec:expan-Green}
Considering a geometry in which a quasi-1D ring
made of a band insulator
is connected to two normal electrodes (see Fig.~\ref{fig:NIN-ext}c),
the fully dressed advanced Green's function $G_{\alpha,\beta}^A$
[see Eqs. (8) and (9) in Ref.~\onlinecite{Caroli}]
\begin{equation}
G_{\alpha,\beta}^A=\frac{g_{\alpha,\beta}^A}
{d_a^A d_b^A-(d'_a)^A(d'_b)^A}
,
\end{equation}
with $d_a^A=1-g_{\alpha,\alpha}^At_{\alpha,a}g_{a,a}^At_{a,\alpha}$
and $d_b^A=1-g_{\beta,\beta}^At_{\beta,b}g_{b,b}^At_{b,\beta}$,
$(d'_a)^A=g^A_{\beta,\alpha}t_{\alpha,a} g_{a,a}^A t_{a,\alpha}$,
$(d'_b)^A=g^A_{\alpha,\beta}t_{\beta,b} g_{b,b}^A t_{b,\beta}$,
is expanded in multiple crossings of the ring:
\begin{eqnarray}
\label{eq:expan}
G_{\alpha,\beta}^A&=&\frac{g_{\alpha,\beta}^A}{d_a^A d_b^A}\\
&+&\frac{(g_{\alpha,\beta}^A)^2 g_{\beta,\alpha}^A}{(d_a^A d_b^A)^2}
\left(t_{\beta,b} g_{b,b}^A t_{b,\beta}\right)
\left(t_{\alpha,a} g_{a,a}^A t_{a,\alpha}\right)
+ ...
\nonumber
\end{eqnarray}
The expansion in Eq.~(\ref{eq:expan})
is justified by the damping of the electron
wave-functions in the insulator\cite{M-PRB,Melin-Feinberg-PRB}:
the second term with three crossings of the insulator
in Eq.~(\ref{eq:expan}) is small compared
to the first term involving a single crossing if we suppose
that half of the perimeter of the ring is large compared to
the coherence length.
Separating propagation from $\alpha$ to $\beta$
along the upper and lower
branches of the ring and including the phase factors related
to the enclosed flux $\Phi$, the combination of
Eq.~(\ref{eq:G1D})-(\ref{eq:expan})
to Eqs.~(\ref{eq:gg0})-(\ref{eq:gg3})
leads to dominant
$h/2e$ oscillations of the conductance with the magnetic flux $\Phi$,
and to a negative magnetoresistance at low field (see
Fig.~\ref{fig:NIN-ext}d). 

More precisely,
the magnetic field is introduced by substituting
Eq.~(\ref{eq:subs})
into Eq.~(\ref{eq:expan}) and keeping the lowest order oscillating
term in the combination $G_{\alpha,\beta}^A G_{\beta,\alpha}^R(\omega)$
entering
the expression of the conductance ${\cal G}(\omega)$
given by Eq.~(\ref{eq:Nozieres}).
Such term contains three advanced and one retarded 
Green's functions, each contributing to a factor $\exp{(i\pi\Phi/2\phi_0)}$.
As a result of averaging over the Fermi phase factors
(see Appendix~\ref{sec:CCNSJ}),
the lowest order oscillating part of the conductance is
proportional to $\cos{(4\pi \Phi/\phi_0)}=
\mbox{Re}[\exp{(4i\pi\Phi/\phi_0)}]$, with $h/2e$-periodic
oscillations
of the conductance with the flux $\Phi$,
typical of a weak localization-like phenomenon.
The sign of the $\cos{(4\pi\Phi/\phi_0)}$
oscillations in the conductance corresponds 
to a negative magnetoconductance
at low field, and it is thus inverted as compared to standard weak
localization in a normal metal (see Fig.~\ref{fig:supp} in the forthcoming
Sec.~\ref{sec:wf}).

Similar features were obtained by Latyshev {\it et al.} \cite{Latyshev}
in experiments on CDWs 
(see the items  3. and 4. in the summary of experiments in 
the Introduction).
The coincidence is explained in Sec.~\ref{sec:phys}
by the 
same underlying weak localization-like subgap tunneling processes 
for band insulators and CDWs.

\begin{figure}
\includegraphics [width=.95 \linewidth]{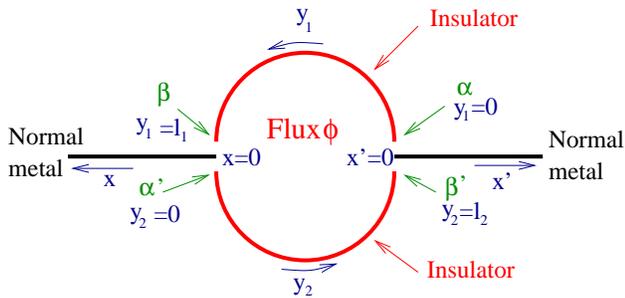}
\caption{(Color online.) Notations used in Sec.~\ref{sec:wf}
for describing an insulating ring connected to two normal electrodes.
\label{fig:supp2}
}
\end{figure}

\subsection{Wave function approach}
\label{sec:wf}
\begin{figure}
\includegraphics [width=.8 \linewidth]{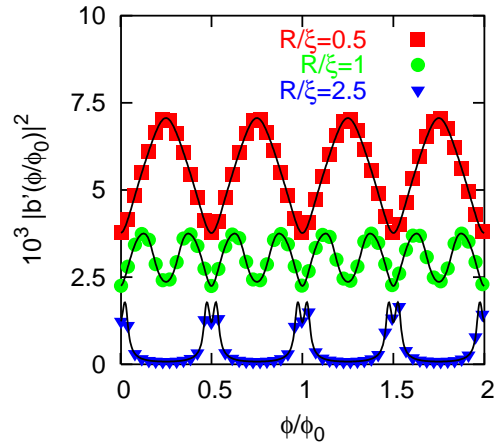}
\caption{(Color online.) Transmission coefficient of the
ring made of a band insulator with the dispersion relation on
Fig.~\ref{fig:disper} as a function of the flux $\Phi$ enclosed in the loop,
normalized to the flux quantum $\phi_0=h/e$. Different curves correspond to
the different values of $R/\xi$ shown on the figure. The value of
$\lambda_F=2\pi/k_F$ is small compared to $\xi$.
\label{fig:supp}
}
\end{figure}

We 
discuss now magneto-oscillations of the tunneling current via wave function
matching \cite{Rammal,Vidal}. The goal is to
confirm qualitatively the Green's function approach
in Sec.~\ref{sec:IIIA}, and to present concrete results for the
magnetic oscillations.
Let us consider the geometry on Fig.~\ref{fig:supp2}
in which a ring (made of an insulator in which wave functions are damped
over a length $\xi$ and oscillate with wave-vector $k_F$)
is connected to two normal electrodes (with propagation of plane waves).
The band insulator is supposed to have the
same BCS dispersion relation
as described above (see Sec.~\ref{sec:ins1}). The corresponding wave-vectors
$\pm k_F \pm i/\xi$ lead to oscillations of the wave-function
with period $\lambda_F=2\pi/k_F$ and to a damping over the insulator
coherence length $\xi$.
Following Refs.~\onlinecite{Rammal,Vidal},
the wave-functions $\psi_L(x)$
in the left normal electrodes, $\psi_1(y_1)$ in the upper
branch of the ring (of length $l_1$),
$\psi_2(y_2)$ in the lower branch (of length $l_2$)
and $\psi_R(x')$ in the right normal metal are given 
by
\begin{widetext}
\begin{eqnarray}
\label{eq:match1}
\psi_L(x)&=&\exp(-ik_F x)+b\exp(ik_F x)\\
\nonumber
\psi_1(y_1)&=& \frac{\psi_\alpha}{\sin{(k_F l_1)}}
\exp{\left(-i\frac{2\pi}{\phi_0}y_1A_0\right)}
\sin{\left(k_F(l_1-y_1)\right)}
\exp{\left(\frac{y_1-l_1}{\xi}\right)}\\
\nonumber
&+&
\frac{\psi_\beta}{\sin{(k_F l_1)}}
\exp{\left(i\frac{2\pi}{\phi_0}(l_1-y_1) A_0\right)}
\sin{\left(k_F y_1\right)} \exp{\left(-\frac{y_1}{\xi}\right)}
\\
\nonumber
\psi_2(y_2)&=& \frac{\psi_{\alpha'}}{\sin{(k_F l_2)}}
\exp{\left(-i\frac{2\pi}{\phi_0}y_2 A_0\right)}
\sin{\left(k_F (l_2-y_2)\right)}
\exp{\left(\frac{y_2-l_2}{\xi}\right)}
\\
\nonumber
&+&
\nonumber
\frac{\psi_{\beta'}}{\sin{(k_F l_2)}}
\exp{\left(i\frac{2\pi}{\phi_0}(l_2-y_2) A_0\right)}
\sin{\left(k_F y_2\right)}
\sin{\left(\frac{-y_2}{\xi}\right)}\\
\psi_R(x')&=&b'\exp(ik_F x')
,
\end{eqnarray}
\end{widetext}
where we choose a radial gauge with a component $A_0$
of the potential vector, and
where the coefficients $b'$ and $b$ correspond to the 
transmission and backscattering amplitudes, and $\psi_\alpha$,
$\psi_\beta$, $\psi_{\alpha'}$ and $\psi_{\beta'}$ denote
the amplitudes of the components of the insulating wave-function localized
on the different nodes. 
Wave-function matching takes the form
$\psi_L(0)=\psi_1(l_1)=\psi_2(0)$, 
$\psi_R(0)=\psi_1(0)=\psi_2(l_2)$, and matching of the derivative
of the wave-function at the left and right nodes is given by
\begin{eqnarray}
\label{eq:der1}
-\frac{\partial\psi_L}{\partial x}(0)
+\frac{\partial \psi_1}{\partial y_1}(l_1)
-\frac{\partial \psi_2}{\partial y_2}(0)&=&0\\
\label{eq:der2}
-\frac{\partial\psi_R}{\partial x}(0)
-\frac{\partial \psi_1}{\partial y_1}(0)
+\frac{\partial \psi_2}{\partial y_2}(l_2)&=&0
,
\end{eqnarray}
as deduced from integrating the Schr\"odinger equation with
respect to coordinates over a small area including a node.
The transmission coefficient
\begin{equation}
\label{eq:T0}
{\cal T}_0(\Phi)=|b'(\Phi)|^2
\end{equation}
is then evaluated by averaging over the Fermi phase factors
$k_F l_1$ and $k_F l_2$, and it is shown on
Fig.~\ref{fig:supp} for different values of $R/\xi$.
The matching equations (\ref{eq:match1}-\ref{eq:der2}) reduce to 
those of a normal metal \cite{Rammal,Vidal}
for $R/\xi\ll 1$ and a
positive magnetoconductance with $\phi_0/2$ periodicity is then
recovered in this range of $R/\xi$ (as for standard weak localization in
a normal metal where increasing a magnetic field suppresses
localization). As seen from Fig.~\ref{fig:supp},
the magnetoconductance is almost $\phi_0/4$-periodic 
at the cross-over
$R\sim\xi$ and becomes again $\phi_0/2$-periodic for $R/\xi\gg 1$,
with a shape of oscillations characteristic of
the evanescent wave-function weak localization-like oscillations
discussed above in Sec.~\ref{sec:IIIA} within microscopic Green's functions.
We obtain $h/2e$-periodicity for $R/\xi\gg 1$,
but not exactly of the form
$\cos{(4\pi\phi/\phi_0)}$ obtained above in Sec.~\ref{sec:expan-Green}
within microscopic Green's functions.
This is because of
a remaining positive magnetoconductance within a small 
low field region around $\phi=n\phi_0/2$ (with $n$
and integer), in agreement with Ref.~\onlinecite{Spivak}.
Higher order harmonics play also a role because of the highly
transparent contacts used for the ring geometry on Fig.~\ref{fig:supp2}
[see Eqs.~(\ref{eq:der1}) and~(\ref{eq:der2})].

\begin{figure}
\includegraphics [width=.8 \linewidth]{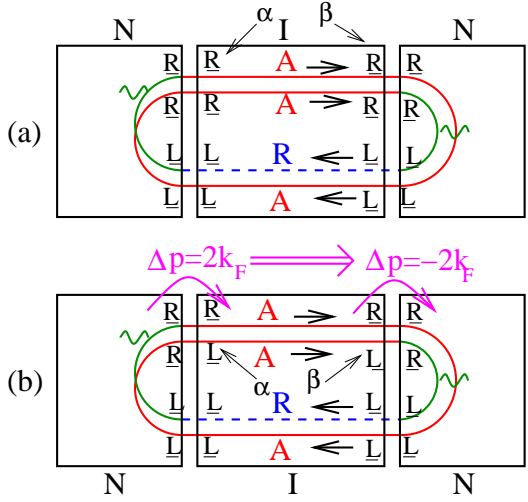}
\caption{(Color online.) Weak localization-like loops 
in a normal metal-insulator-normal metal junction
with two sets of right-left labels. The right-left branches and
therefore the momentum of the tunneling electrons is conserved
in (a). 
The conductance
is vanishingly small for (a) because of the absence of propagation 
from $\alpha$ to $\beta$ of the advanced-advanced transmission
mode [$\overline{g_{\alpha,\beta,
\underline{R},\underline{R}}^A 
g_{\alpha,\beta,\underline{R},\underline{R}}^A}\simeq 0$ 
--
see Eq.~(\ref{eq:g4x4})].
(b) shows a process leading to a finite tunneling current
[$\overline{g_{\alpha,\beta,
\underline{R},\underline{R}}^A 
g_{\alpha,\beta,\underline{L},\underline{L}}^A}$ is
limited by the insulator coherence length --
see Eq.~(\ref{eq:g4x4})]. Transfers of momentum by
$\Delta p =\pm 2 k_F$ propagate across the
insulator according to the arrows, in parallel 
to evanescent wave charge tunneling.
\label{fig:rightleft}
}
\end{figure}
\subsection{Coupling to a momentum channel}
\label{sec:recoil}
\label{sec:q1D}
Now, we note that propagation across the insulator 
defines a ``tunnel'' of cross section area $\xi_0^2$,
with $\xi_0\sim a_0\epsilon_F/\Delta$ the coherence length
($a_0$ is the lattice spacing and $\epsilon_F$ the Fermi energy).
Such a narrow channel is compatible with 
the following additional assumption:
the insulator (with the dispersion relation on
Fig.~\ref{fig:disper}) consists of linear
1D chains perpendicular to the interfaces.
Electrons on the right and left branches
(denoted by $\underline{R}$ and $\underline{L}$) 
are taken into account
according to
Sec.~\ref{sec:ins1}. We deduce that 
the specific set of $\underline{R}$ and $\underline{L}$ labels on
Fig.~\ref{fig:rightleft}a 
does not contribute to the conductance once
the summation over the Fermi oscillations in different channels is
carried out. By contrast, branch crossing at the interfaces
(see the $\underline{R}$ and $\underline{L}$ labels on
Fig.~\ref{fig:rightleft}b)
contribute for a finite value to the conductance,
and involve a
transfer of $2 k_F$ momentum from one interface to
the other across the
insulating electrode. In a quasi-1D geometry,
electrons above the gap on the right branch (label $\underline{R}$)
of the BCS-like dispersion relation
can propagate physically to the left or to the right according
to their group velocity (see Fig.~\ref{fig:disper}).
This means that a 
recoil of the insulator is a necessary condition for weak localization-like
subgap tunneling with quasi-1D insulators in the geometry of
Fig.~\ref{fig:rightleft}. Moreover, transfers of $2 k_F$ momentum
are the hall-mark of CDW Andreev processes
\cite{Kasatkin,Artemenko-Andreev,AR}, which suggests a
connection 
between the special type of tunnel junctions considered here and
the CDW case (see Sec.~\ref{sec:phys}). 

Finally, disorder in the normal electrode tends to
localize the electron wave functions in the vicinity of the
interfaces. The processes on Fig.~\ref{fig:rightleft} are
thus facilitated by diffusive motion in the normal electrodes.

\begin{figure*}
\includegraphics [width=1. \linewidth]{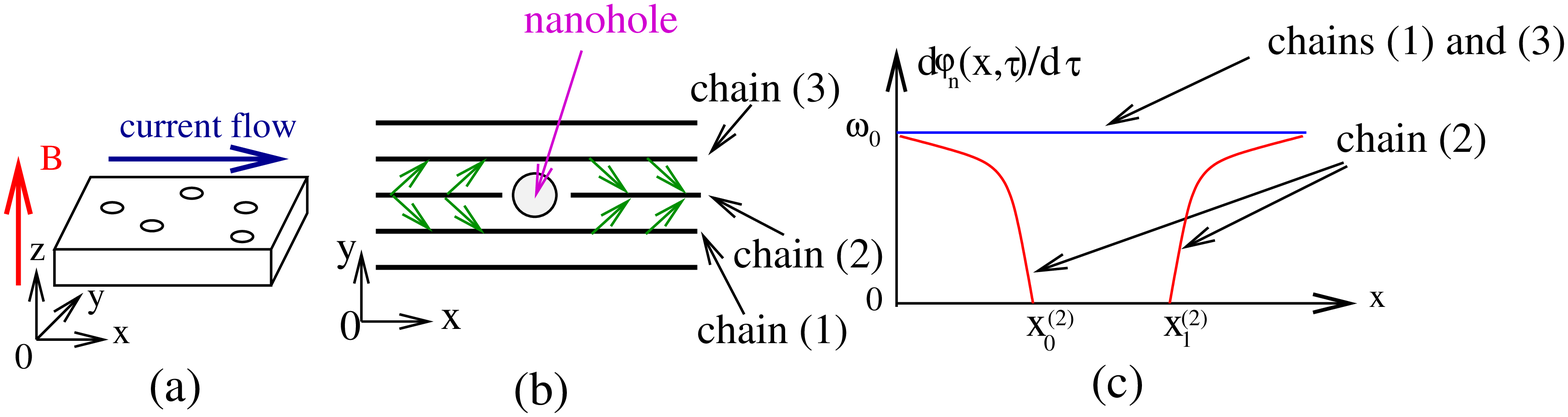}
\caption{(Color online.) Schematic representation of (a) a film
of NbSe$_3$ pierced by nanoholes with a magnetic field ${\bf B}$
along the $z$ axis. The current is supposed to flow on average
along the $x$ axis parallel to the chains.
(b) shows a nanohole
interrupting a single CDW chain along the $x$ axis.
The arrows on (b)
represent schematically the emission and absorption
of normal carriers due
to the slowing down and acceleration
of the sliding motion at the left
and right of the nanohole respectively.
The profile
of $\partial \varphi_n(x,\tau)/\partial \tau$
along chains $n=1,2,3$ is shown schematically on (c).
\label{fig:schema}
}
\end{figure*}
\begin{figure}
\includegraphics [width=1. \linewidth]{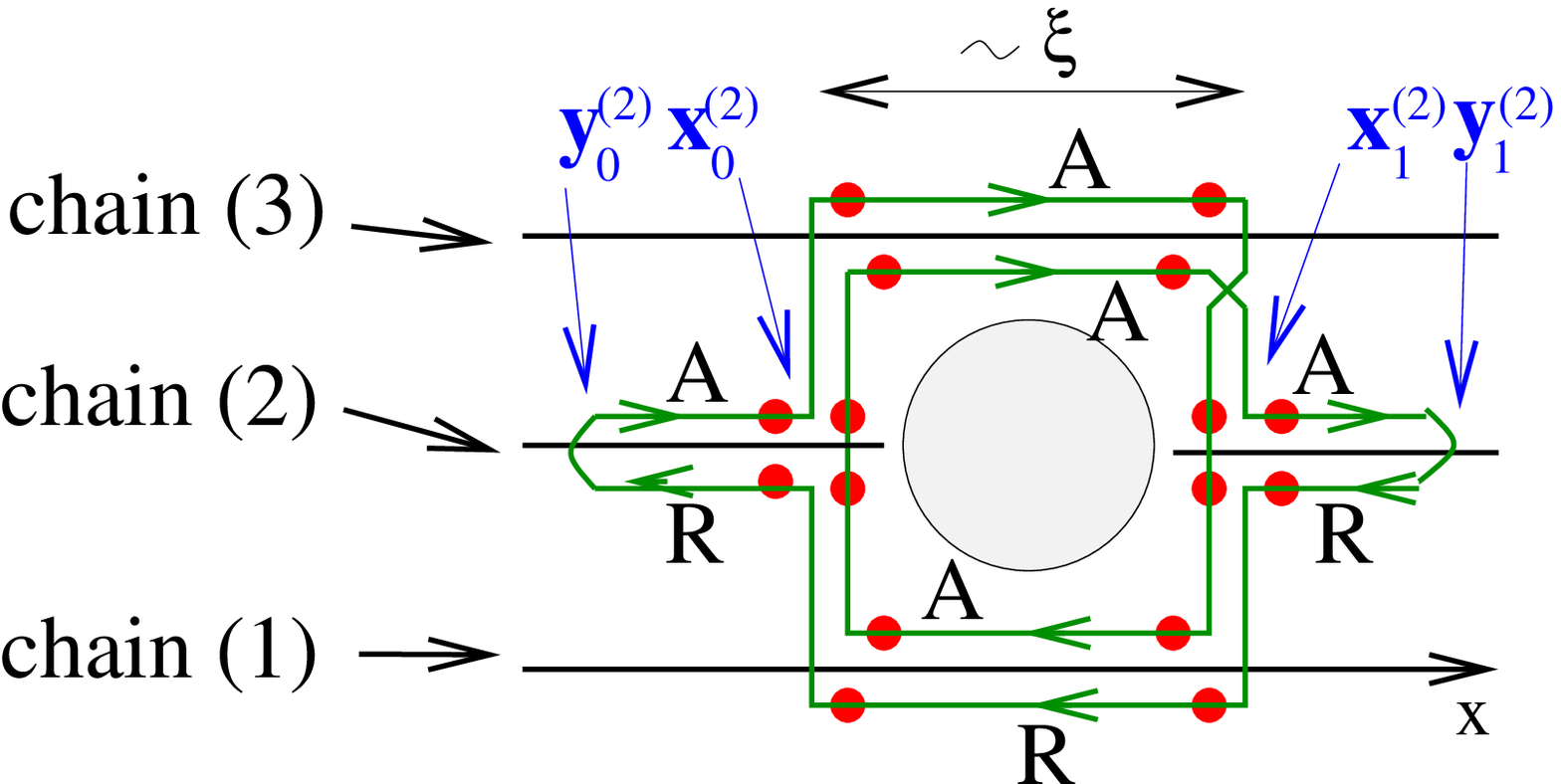}
\caption{(Color online.) The same weak localization-like tunneling
process as on Fig.~\ref{fig:loop}c for a nanohole in a CDW,
as on Fig.~\ref{fig:schema}.
\label{fig:schemabis}
}
\end{figure}

\section{Subgap tunneling by a quantum interference effect
around a nanohole in a charge density wave}
\label{sec:phys}

Now, we present our main result and show that 
weak localization-like subgap
tunneling discussed above for tunnel junctions
(Sec.~\ref{sec:jonctiontunnel}) 
leads to a quantum
interference effect also in charge density waves, therefore explaining
the experiment by Latyshev {\it et al.}\cite{Latyshev} on oscillations of
the CDW current around a nanohole.
Before discussing weak localization-like subgap tunneling in
Sec.~\ref{sec:TOTO2}, we provide in Sec.~\ref{sec:TOTO1} a mechanism
of transport around a nanohole in a CDW.
The disordered
case is discussed in Appendix~\ref{app:des}.
\begin{figure}
\includegraphics [width=.6 \linewidth]{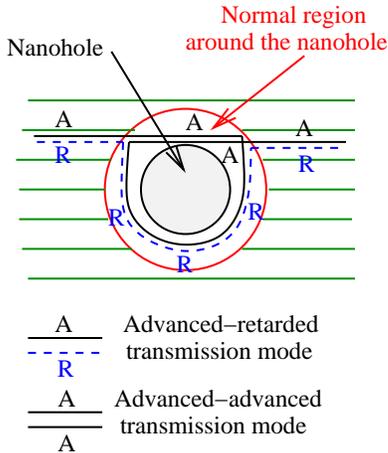}
\caption{(Color online.) Schematic representation of a
weak localization-like tunneling events involving an ``advanced-advanced''
transmission mode, and an ``advanced-retarded''
transmission mode in the presence of a normal region around the nanohole,
at energies smaller than the island level spacing.
\label{fig:composite}
}
\end{figure}

\subsection{Coupling between the sliding motion and weak localization-like
subgap tunneling}
\label{sec:TOTO1}
\subsubsection{Notations}
The mechanism discussed now for CDW transport, based on weak localization-like
subgap tunneling around a nanohole,
is first discussed for
a few interrupted chains coupled by a transverse hopping $t_\perp$.
The more realistic case of a normal island around the nanohole will
be discussed afterwards.
From the point of view of notations,
the interrupted chains are
labeled by $k=2,...,N-1$, and are
connected by transverse hopping terms to the two uninterrupted chains
$k=1$ on top and $k=N$ on bottom (see Fig.~\ref{fig:schema} for $N=3$
and Fig.~\ref{fig:loop}c for $N=5$).
Chains $k=2,...,N-1$ are interrupted at positions $x_0^{(k)}$
at the left of the nanohole, and at $x_1^{(k)}$ at
the right of the nanohole
(see $x_0^{(2)}$ and $x_1^{(2)}$
on Figs.~\ref{fig:schema}b and~\ref{fig:schemabis} for a single
interrupted chain).

\subsubsection{Deceleration and acceleration of the 
sliding motion in interrupted chains}

Assuming an overall sliding motion,
the phase $\varphi_k(x,\tau)$
at position $x$ along chains $k=1$ and $k=N$ and time $\tau$
follows $\partial \varphi_k(x,\tau)/\partial \tau=\omega_0$,
with $\omega_0$ the sliding frequency, corresponding to the
relation (see for instance Ref. \onlinecite{ps13})
\begin{equation}
j^{(k)}(x,\tau)=\frac{e}{\pi} \frac{\partial \varphi_k(x,\tau)}
{\partial \tau}
\end{equation}
between the CDW current and the time derivative
of the CDW phase variable in chain $k$.
The
boundary condition on the hole leads to 
$\partial \varphi_k(x,\tau)/
\partial \tau=0$ at $x=x_i^{(k)}$ ($i=0,1$ and $k=2,...,N-1$)
(no collective current is flowing
across $x_0^{(k)}$ and $x_1^{(k)}$ in the direction parallel to the chains),
and to $\partial \varphi_k(x,\tau)/\partial \tau=\omega_0$ for
$x\ll x_0^{(k)}$ and $x\gg x_1^{(k)}$ 
(the collective sliding motion is recovered
far away from the nanohole).
The resulting profile 
of $\partial \varphi_k(x,\tau)/\partial \tau$
(see Fig.~\ref{fig:schema}c for $N=3$)
corresponds to the conversion of the CDW current into
a normal current \cite{B1,B2,B3,B4}
$\partial \rho_k(x,\tau)/\partial \tau$,
emitted according to the arrows on Fig.~\ref{fig:schema}b,
from the intermediate chain labeled by $k$:
\begin{equation}
\frac{\partial \rho_k(x_0^{(k)},\tau)}{\partial \tau}
=-{e\over \pi}\frac{\partial^2
\varphi_k(x_0^{(k)},\tau)}{\partial x \partial \tau}
,
\end{equation}
as deduced  from the continuity equation.

\subsubsection{Charge accumulation}
\label{sec:smallness}
Quasiparticles emitted from the slowing down of the
sliding motion at the left of the nanohole
are reabsorbed at its right where the sliding motion
accelerates, leading to charge accumulation at the
extremities of the interrupted chains at the left of the hole,
described by the chemical potential $\delta \mu \agt \Delta$.
The current $I_k(\Phi)$ emitted from chain $k$ is given by
\begin{equation}
\label{eq:deltamu}
I_k(\Phi)=\frac{e}{h} \sum_m
\int_0^{\delta\mu} {\cal T}_{k\rightarrow m}(\Phi,t_\perp,\hbar\omega)
d(\hbar\omega)
,
\end{equation}
where ${\cal T}_{k\rightarrow m}(\Phi,t_\perp,\hbar \omega)$
is the total dimensionless transmission coefficient at energy 
$\hbar \omega$ transfering electrons from chain $k$
at the left of the nanohole to chain $m$ at its right
(see chains $k$ and $m$ on Fig.~\ref{fig:loop}).
The transmission coefficient ${\cal T}_{k\rightarrow m}$ in
Eq.~(\ref{eq:deltamu}) is a multichannel generalization of
${\cal T}_0$ in Eq.~(\ref{eq:T0}). Similarly to the
case of band insulators (see Sec.~\ref{sec:PRELIM}),
the subgap tunneling current for the processes on
Fig.~\ref{fig:schemabis}
is $h/2e$-periodic
as a function of the enclosed flux $\Phi$
[see Eq.~(\ref{eq:Trans}) below obtained from Eq.~(\ref{eq:subs})].
The oscillations appear only in the presence of the sliding
motion (item 2. in the Introduction).
The smallness of interchain couplings in the transmission coefficient
can be balanced 
by the integral over energy in Eq.~(\ref{eq:deltamu}),
up to the large value of the
Peierls gap in the compound NbSe$_3$ used 
by Latyshev {\it et al.}\cite{Latyshev}.
The non modulated part of the current for a single
interrupted chain is proportional to $I_0\sim(e/h)
(t_\perp/T)^4(\delta\mu-\Delta)$ while the modulated
part $I_{mod}=(e/h)(t_\perp/T)^8\Delta$ is independent on
$\delta\mu-\Delta$. The ratio $I_{mod}/I_0=
(t_\perp/T)^4\Delta/(\delta \mu-\Delta)$ 
is thus a function of the values of
$t_\perp/T$ and of $(\delta\mu-\Delta)/\Delta$.

However,
it was already noticed \cite{Visscher} that 
the damaged region of the CDW around a nanohole 
is likely to be normal,
in which case the relative amplitude of the modulation may
be large, as suggested by Fig.~\ref{fig:supp} for a single channel.
The corresponding tunneling process is shown on
Fig.~\ref{fig:composite} at energy smaller
than the normal region level spacing $\delta$. 
Propagation across
the normal region supports ``advanced-advanced''
transmission modes because the level spacing $\delta$
plays the role of a gap. 
The value of $\delta$ is comparable to the Peierls gap,
as it can be seen from the estimate $\hbar v_F/D$,
with $D$ the diameter of the nanohole, leading to
the rough estimate
$\delta/k_B \sim 100\,$K,
with $v_F\simeq10^5\,$ms$^{-1}$ as an order of magnitude
of the Fermi velocity. As expected, the level spacing of
an object of size $D\sim\xi$ is comparable to the Peierls
gap $\Delta$.

\subsection{Weak localization-like subgap tunneling transmission
coefficient}
\label{sec:wl-diag}
\label{sec:TOTO2}

We evaluate now the 
weak localization-like subgap tunneling transmission coefficient
given by the diagram on Fig.~\ref{fig:schemabis}, in the absence of
normal region around the nanohole as on Fig.~\ref{fig:composite}.
It takes the form
\begin{equation}
\label{eq:Trans}
{\cal T}
\left(\Phi,t_\perp,\hbar\omega\right)=
\left(\frac{t_\perp}{T}\right)^8
{\cal F}\left(\hbar\omega\right)
\Xi(\hbar\omega) \cos{\left(\frac{2\Phi}{\phi_0}\right)}
,
\end{equation}
The factor $\Xi(\hbar\omega)$ encodes the damping of subgap transmission:
\begin{equation}
\Xi(\hbar\omega) = \exp{\left(-\frac{y_1^{(2)}-y_0^{(2)}}{\xi(\omega)}\right)}
\exp{\left(-\frac{x_1^{(2)}-x_0^{(2)}}{\xi(\omega)}\right)}
,
\end{equation}
and ${\cal F}\left(\hbar\omega\right)$
takes the form
\begin{eqnarray}
\label{eq:F-generic}
&&{\cal F}\left(\hbar \omega\right)=
\mbox{Re} \left[
4 \left|g_0(\omega)\right|^4 \left\{(g_0(\omega))^2+
(f_0(\omega))^2\right\}
\right.\\
&&\times \left.
\left\{
\left|g_0(\omega)\right|^2+\left|f_0(\omega)\right|^2\right)\right\}
,
\nonumber
\end{eqnarray}
with $g_0(\omega)$ and $f_0(\omega)$ given by Eqs.~(\ref{eq:g0})
and (\ref{eq:f0}). The term $\left|g_0(\omega)\right|^2+
\left|f_0(\omega)\right|^2$ corresponds to the ``advanced-retarded''
transmission mode in the lower branch (see Fig.~\ref{fig:schema}d)
and the term $(g_0(\omega))^2+
(f_0(\omega))^2$ accounts for the ``advanced-advanced'' transmission
mode in the upper branch. The later involves
transmission of momentum in parallel to evanescent wave tunneling,
similarly to the tunnel junction discussed in Sec.~\ref{sec:jonctiontunnel}
(see Fig.~\ref{fig:rightleft}b).
The coherence length $\xi(\omega)$ at energy $\hbar \omega$
is given by $\xi(\omega)=\hbar v_F/\sqrt{|\Delta|^2-(\hbar \omega)^2}$,
with $v_F$ the Fermi velocity.

The sliding motion induces a time dephasing of weak localization-like
tunneling because the CDW phase evolves in time in the course
of weak localization winding.
Such couplings are however not probed in experiments such
as in Ref.~\onlinecite{Latyshev} because the sliding motion
is slow compared to the time scale $\hbar /\Delta$, with
$\Delta$ the Peierls gap.

A ballistic evaluation of the transmission coefficient
in the CDW chains
is justified by the fact that tunneling quasiparticles travel over
extremely short
distances corresponding to the diameter of the hole,
comparable to the CDW coherence length of order $10\,$nm
in experiments \cite{Latyshev}.
Similar results were obtained in a different limit by
treating disorder
in the ladder approximation along Ref.~\onlinecite{Smith}
(the principle of the calculation is detailed in
Appendix~\ref{app:ladder}).

\section{Conclusions}
\label{sec:conclu}
To conclude, we discussed for CDWs a mechanism of tunneling via
quantum interference effect initially proposed for
superconducting hybrids \cite{M-PRB,DM,Altland}. 
The considered tunneling mechanism combined to charge accumulation
due to the deceleration of the CDW at the approach of the nanohole
leads to the same features as in the experiment by Latyshev
{\it et al.}\cite{Latyshev} on $h/2e$ oscillations of the
CDW current around a nanohole (see the Introduction):
1. Weak localization-like loops
due to higher order terms in the tunnel amplitudes are present
even without nanoholes.
They induce no oscillations in the magnetoresistance 
in the absence of nanohole because of the absence of charge
accumulation in this case;
2. No charge accumulation is
present with nanoholes but without sliding motion;
3. $h/2e$-periodic oscillations of the resistance as a function
of the magnetic flux are obtained with nanoholes and with sliding
motion because the diameter of the nanohole is comparable to the
coherence length, so that weak localization-like loops enclose
approximately the same area as the nanohole; 4. The
positive magnetoresistance at low field is already obtained
for the normal tunnel junction;
5. 
The only energy/temperature scale in weak localization-like
subgap tunneling is the gap or 
the level spacing of the metallic island, comparable
in magnitude to the Peierls gap.

An underlying issue is whether the experiment
by Latyshev {\it et al.}\cite{Latyshev} provides evidence 
for an interference effect associated to
the collective quantum mechanical CDW ground state.
The collective momentum channel 
may be realized by a recoil in the specific case of a
quasi-1D insulator.
The analogy between CDWs and quasi-1D insulators
and on the other hand the expected normal island 
in the Latyshev {\it et al.} experiment \cite{Latyshev} shows
that propagation through the CDW condensate is not a necessary
condition for the modulations of the resistance as a function
of a magnetic field.

\section*{Acknowledgments}
R.M. thanks Yu. Latyshev, P. Monceau and A.A. Sinchenko for 
stimulating discussions on their experiments,
S. Brazovski for a crucial discussion 
at the early stages of this work, and acknowledges
a fruitful discussion with D. Carpentier and E. Orignac,
as well as with J. Dumas and with B. Dou\c{c}ot.
S. D. and R. M. thank D. Feinberg, S. Florens, M. Houzet 
and R. Whitney for having provided
useful insights and suggestions.
Ref.~\onlinecite{Altland} was provided 
to us by M.V. Feigelman.

\appendix

\begin{figure*}
\includegraphics [width=.7 \linewidth]{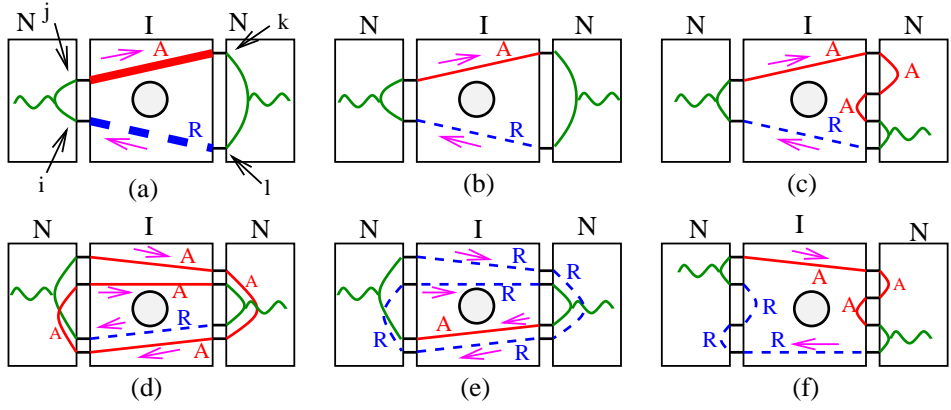}
\caption{(Color online.) Top view of Fig.~\ref{fig:NIN-ext}a showing the
first terms in perturbation theory in the tunnel amplitude
for a normal metal-insulator-normal metal (NIN) tunnel junction.
The exact transport formula is shown schematically
on (a), where the thick solid (dashed) lines correspond
to the fully dressed advanced (retarded) Green's functions. 
The wavy lines indicate the insertion of the density
of states in the normal electrodes (see Fig.~\ref{fig:NIN-ext}b).
The other panels
correspond to a few of the first
low order terms in the perturbative expansion in
the tunnel amplitudes. As for usual weak localization in a normal
metal (see Fig.~\ref{fig:loop}a), the 
weak localization-like tunneling processes on
(d) and (e) lead to dominant $h/2e$ oscillations as a function
of the magnetic flux once the summation over all channels
has been carried out. 
\label{fig:NIN}
}
\end{figure*}

\section{Tunneling through an insulator with planar interfaces}
\label{sec:CCNSJ}

\subsection{Multichannel transport formula}
The exact
generalization of the 1D transport formula
(see Eq.~(\ref{eq:Nozieres})) 
to a multichannel N$_a$IN$_b$ tunnel junction with extended interfaces
(see Fig.~\ref{fig:NIN-ext}a)
is given by
\begin{eqnarray}
\nonumber
{\cal G}(\omega)= \sum_{i,j,k,l} &&
4\pi^2 \frac{e^2}{h} 
\rho_{a_i,a_j}(\omega) t_{a_j,\alpha_j} 
G_{\alpha_j,\beta_k}^A(\omega) t_{\beta_k,b_k}
\\
&\times& 
\rho_{b_k,b_l}(\omega)
t_{b_l,\beta_l} G_{\beta_l,\alpha_i}^R(\omega)
t_{\alpha_i,a_i}
\label{eq:G-multi}
,
\end{eqnarray}
where 
an underlying tight-binding lattice is assumed.
The density of states $\rho_{a_i,a_j}$ connects
the two sites $a_i$ and $a_j$ at the same interface.
The summations over $(i, j)$ and $(k,l)$ run over all sites at the
N$_a$I and IN$_b$ interfaces respectively.
The transport formula given by Eq.~(\ref{eq:G-multi}) is 
represented on
Fig.~\ref{fig:NIN}a, where the bold red and dashed blue
lines correspond to $G_{\alpha,\beta}^A$ and $G_{\beta,\alpha}^R$
respectively. The densities of states in the normal electrodes
are represented by the connection of a wavy line
on the figure.

\subsection{Perturbative expansion}
\label{sec:IIB}
Higher order tunnel processes containing weak localization-like
contributions are obtained by
expanding systematically Eq.~(\ref{eq:G-multi}) in the tunnel amplitudes
according to the Dyson equations.
Fig.~\ref{fig:NIN} b-f shows combinations of some terms in
$G_{\alpha,\beta}^A$ to some terms in $G_{\beta,\alpha}^R$
The weak localization-like diagrams on Fig.~\ref{fig:NIN}d
and e are made of three ``advanced'' and one ``retarded''
Green's functions, and look similar to
the diagram
on Fig.~\ref{fig:loop}b corresponding to a disorder
self-energy in the bulk of a superconductor.

\subsection{Averaging over the conduction channels}
\label{sec:IIC}
The diagrams appearing in perturbation theory
are averaged over the different
``channels'' in real space, corresponding to a summation over the
discrete tight-binding
sites at which diagrams cross the interfaces. The result of
the channel averaging procedure depends on assumptions about the
insulator band structure.

Averaging over the conduction channels amounts to evaluating
the integrals
\begin{eqnarray}
\label{eq:gg0}
&&\overline{g^A(R_{\alpha,\beta},\omega)g^R(R_{\alpha,\beta},\omega)}\\
\label{eq:gg1}
&=&
\frac{k_F}{2\pi} \int_{R_{\alpha,\beta}}^{R_{\alpha,\beta}+2\pi/k_F}
g^A(r,\omega)g^R(r,\omega) dr \\
\label{eq:gg2}
&=&
{1\over 2} \left(\frac{\pi \rho_N}{k_F R_{\alpha,\beta}}\right)^2
\frac{(\hbar \omega)^2}{|\Delta|^2-(\hbar\omega)^2}
\exp{\left(-\frac{2R_{\alpha,\beta}}{\xi(\omega)}\right)}\\
&=& \overline{g^A(R_{\alpha,\beta},\omega)g^A(R_{\alpha,\beta},\omega)} 
\label{eq:gg3}
,
\end{eqnarray}
where $g^A(R_{\alpha,\beta},\omega)$ is given by Eq.~(\ref{eq:gA-ins}),
and where the last equality is valid for $\hbar \omega<\Delta$ below the gap.
To summarize, Aharonov-Bohm like oscillations are washed out by
channel averaging with the band structure on Fig.~\ref{fig:band} and
$h/2e$-periodic weak localization-like diagrams
on Figs.~\ref{fig:NIN}d and e contribute to leading order (in the
tunnel amplitudes) to the
oscillations of the conductance as a function of the magnetic flux.
The same conclusion holds for the ring geometry in 
Sec.~\ref{sec:expan-Green}.

\begin{figure}
\includegraphics [width=.8 \linewidth]{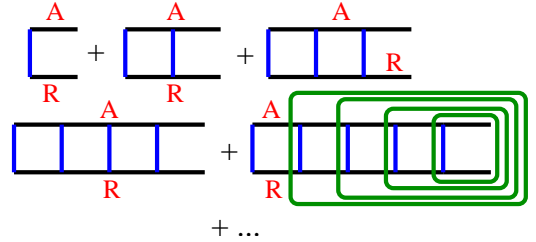}
\caption{(Color online.) The ladder series for
the transmission coefficient. The vertical blue lines
correspond to impurities in the ladder approximation.
The green boxes corresponding
to Eq.~(\ref{eq:Ftaun}) indicate the
recursive elimination of the impurities from one extremity
of the ladder \cite{Smith}.
\label{fig:salpeter}
}
\end{figure}
\section{Evaluation of the transmission coefficient of a disordered CDW}
\label{app:ladder}
\label{app:des}
\subsection{Green's function of a disordered CDW}
The Green's function \cite{Gorkov,ictp} of a disordered CDW is evaluated
in the Born approximation 
as in the superconducting case, where we introduce
forward- and backward-scattering potentials $u$ and $v$.
With the notation in Ref. \onlinecite{Smith} in the superconducting case,
we find
\begin{equation}
\hat{G}(\xi,\omega)=\frac{\hbar\overline{\omega}+
\overline{\xi} \hat{\tau}_3
+\overline{\Delta} \hat{\tau}_1}{(\hbar\overline{\omega})^2
-\overline{\xi}^2-\overline{\Delta}^2}
,
\end{equation}
with $\hat{\tau}_1$ and $\hat{\tau}_3$ the Pauli matrices given below,
with $\overline{\omega}=\omega(1+\alpha)$,
$\overline{\xi}=\xi+\beta$ and $\overline{\Delta}=\Delta(1+\gamma)$,
with
\begin{eqnarray}
\alpha &=& \frac{1}{\tau_f\sqrt{\Delta^2-\omega^2}}
+\frac{1}{\tau_b\sqrt{\Delta^2-\omega^2}}\\
\nonumber
&=&-\left(\left|u\right|^2+\left|v\right|^2\right)
\int \lambda_F \frac{dk}{2\pi}
\frac{1}{(\hbar\omega)^2-\Delta^2-\xi^2}\\
\beta
&=&\left(\left|u\right|^2+\left|v\right|^2\right)
\int \lambda_F \frac{dk}{2\pi}
\frac{\xi}{(\hbar\omega)^2-\Delta^2-\xi^2}\\
\gamma &=& \frac{1}{\tau_f\sqrt{\Delta^2-(\hbar\omega)^2}}\\
&=&\left|u\right|^2
\int \lambda_F \frac{dk}{2\pi}
\frac{1}{(\hbar\omega)^2-\Delta^2-\xi^2}
\nonumber
,
\end{eqnarray}
where $\lambda_F$ is the Fermi wave-vector,
$\hbar\omega$ the energy, $\Delta$ the
CDW gap and $\xi$ the kinetic energy with respect
to the Fermi level. The notations $\tau_f$ and $\tau_b$ 
stand for the
forward and backward scattering times respectively.
We keep in the following calculations
a finite shift $\beta$ of the chemical potential
that does however not enter the properties that we consider.

\subsection{Evaluation of ladder diagrams for the transmission coefficient}
To include disorder\cite{Vanyolos}, the
transmission coefficients in the ladder approximation 
(see Fig.~\ref{fig:salpeter}) are evaluated 
according to Smith and Ambegaokar\cite{Smith}
by iterations of 
\begin{eqnarray}
\label{eq:Ftaun}
&&\hat{F}(\hat{\tau}_n)=\\
\nonumber
&&\int \frac{dk}{2\pi}
\overline{
\left(\begin{array}{cc} u & w \\ v & u \end{array} \right)^\dagger
\hat{G}^{A}(k,\omega) \hat{\tau}_n \hat{G}^{R}(k+q,\omega)
\left(\begin{array}{cc}
u & w \\
v & u \end{array} \right)}
,
\end{eqnarray}
where $u$, $v$ and $w$ are Gaussian distributed random variable. 
The Pauli matrices $\hat{\tau}_n$ are such that
\begin{eqnarray}
\hat{\tau}_0=\left(\begin{array}{cc} 1 & 0 \\ 0 & 1 \end{array} \right)
\mbox{ , }
\hat{\tau}_1=\left(\begin{array}{cc} 0 & 1 \\ 1 & 0 \end{array} \right)\\
i\hat{\tau}_2=\left(\begin{array}{cc} 0 & 1 \\ -1 & 0 \end{array} \right)
\mbox{ , }
\hat{\tau}_3=\left(\begin{array}{cc} 1 & 0 \\ 0 & -1 \end{array} \right)
.
\end{eqnarray}

We find
\begin{eqnarray}
\hat{F}(\hat{\tau}_0)&=&(-A+B)(1-\alpha) \hbar \omega \hat{\tau}_3\\
\nonumber
&&-3A(1+\gamma)\Delta i\hat{\tau}_2 \\
\hat{F}(\hat{\tau}_1)&=&(A-B)(1+\gamma)\Delta \hat{\tau}_3\\
\nonumber
&&-3A(1-\alpha)\hbar\omega i\hat{\tau}_2 \\
\hat{F}(i\hat{\tau}_2)&=&3(A+B)(1+\gamma) \Delta \hat{\tau}_0\\
\nonumber
&&-3A(1-\alpha)\hbar\omega \hat{\tau}_1 \\
\hat{F}(\hat{\tau}_3)&=&(-A-B)(1-\alpha)\hbar\omega \hat{\tau}_0\\
\nonumber
&&-A(1+\gamma)\Delta \hat{\tau}_1 
,
\end{eqnarray}
with
\begin{eqnarray}
A&=&\frac{m^2}{4\hbar^4 k_F^5}
\frac{(8k_F^2+19q^2) \overline{|u|^2}}
{\left(\Delta^2(1+\gamma)^2-\hbar\omega^2(1-\alpha)^2\right)^{1/2}}\\
B&=&\frac{m^2}{4\hbar^4 k_F^5}
\frac{(8k_F^2+19q^2) \overline{|v|^2}}
{\left(\Delta^2(1+\gamma)^2-\hbar\omega^2(1-\alpha)^2\right)^{1/2}}\\
.
\end{eqnarray}
Acting twice with $\hat{F}$ according to $\hat{G}=\hat{F}^2$ leads to
the closed $2\times 2$ equations
\begin{eqnarray}
\hat{G} \left(\begin{array}{c} \hat{\tau}_0\\\hat{\tau}_1
\end{array} \right)
&=& \left(\begin{array}{cc}
a & b \\ b' & a' \end{array} \right)
\left(\begin{array}{c} \hat{\tau}_0 \\ \hat{\tau}_1
\end{array} \right)\\
\hat{G} \left(\begin{array}{c} i\hat{\tau}_2\\\hat{\tau}_3
\end{array} \right)
&=& \left(\begin{array}{cc}
c & d \\ d' & c' \end{array} \right)
\left(\begin{array}{c} i\hat{\tau}_2 \\ \hat{\tau}_3
\end{array} \right)
,
\end{eqnarray}
with
\begin{eqnarray}
a&=&(A^2-B^2)(1-\alpha)^2(\hbar\omega)^2\\
\nonumber
&&-9A(A+B)(1+\gamma)^2\Delta^2\\
b&=&A(10 A-B)(1-\alpha)(1+\gamma) \hbar\omega\Delta\\
b'&=&-(A+B)(10A-B)
(1-\alpha)(1+\gamma)\\
\nonumber
&&\times \hbar\omega\Delta\\
a'&=&9A^2(1-\alpha)^2(\hbar\omega)^2\\
\nonumber
&&-A(A-B)(1+\gamma)^2\Delta^2
,
\end{eqnarray}
and expressions of the same type for $c$, $d$, $c'$, $d'$:
\begin{eqnarray}
c&=&-3(A-B)(2A+B)(1-\alpha)(1+\gamma)\\
\nonumber
&&\times \hbar\omega\Delta\\
d&=&-9A\left[(A+B)(1+\gamma)^2\Delta^2\right.\\
\nonumber
&&\left.-A(1-\alpha)^2(\hbar\omega)^2
\right]\\
c'&=&(A-B)\left[(A+B)(1-\alpha)^2(\hbar\omega)^2\right.\\
\nonumber
&&-\left.A(1+\gamma)^2\Delta^2\right]\\
d'&=&3A(2A+B)(1-\alpha)(1+\gamma)\hbar\omega\Delta
.
\end{eqnarray}
The final step is to decompose the initial condition
on the eigenvectors of $\hat{G}$ and evaluate the coefficients
in matrix geometric series such as
\begin{eqnarray}
\sum_{n=1}^{+\infty}
\hat{G}^n \left(\begin{array}{c} \hat{\tau}_0\\0\end{array}\right)
&=&A_1 \left(\begin{array}{c} \hat{\tau}_0 \\ 0 \end{array} \right)
+B_1 \left(\begin{array}{c} 0 \\ \hat{\tau}_1 \end{array} \right)\\
\sum_{n=1}^{+\infty} 
\hat{G}^n \left(\begin{array}{c} 0\\\hat{\tau}_1\end{array}\right)
&=&A_2 \left(\begin{array}{c} \hat{\tau}_0 \\ 0 \end{array} \right)
+B_2 \left(\begin{array}{c} 0 \\ \hat{\tau}_1 \end{array} \right)
\end{eqnarray}
We find
\begin{eqnarray}
A_1 &=& X
\frac{\lambda_+}{1-\lambda_+} \psi_+^{(1)}
+ Y \frac{\lambda_-}{1-\lambda_-} \psi_-^{(1)}\\
B_1 &=& X
\frac{\lambda_+}{1-\lambda_+} \psi_+^{(2)}
+ Y \frac{\lambda_-}{1-\lambda_-} \psi_-^{(2)}\\
A_2 &=& -X'
\frac{\lambda_+}{1-\lambda_+} \psi_+^{(1)}
+ X' \frac{\lambda_-}{1-\lambda_-} \psi_-^{(1)}\\
B_2 &=& -X'
\frac{\lambda_+}{1-\lambda_+} \psi_+^{(2)}
+ X' \frac{\lambda_-}{1-\lambda_-} \psi_-^{(2)}
,
\end{eqnarray}
with
\begin{eqnarray}
X&=&{1\over 2}\left(1-\frac{b'-a}{(b'-a)^2+4a'b}\right)\\
Y&=&{1\over 2}\left(1+\frac{b'-a}{(b'-a)^2+4a'b}\right)\\
X'&=&\frac{b}{(b'-a)^2+4a'b}\\
\lambda_\pm&=&{1\over 2}\left(a+b'\pm\sqrt{(a-b')^2+4a'b}\right)\\
\psi_\pm^{(2)}&=&\frac{b'-b\pm\sqrt{(a-b')^2+4a'b}}{2b}
\end{eqnarray}
and $\psi_\pm^{(1)}=1$.
Carrying out the same calculation in the sector
$(i\hat{\tau}_2,\hat{\tau}_3)$ and
evaluating geometric series like
\begin{equation}
\sum_{n=1}^{+\infty}
\hat{G}^n \hat{F}
\left(\begin{array}{c} \hat{\tau}_0\\0\end{array}\right)
\end{equation}
in both sectors
leads to an expression of all right-left components of the
transmission coefficient of a disordered CDW. The later
can be used to evaluate the weak localization-like
subgap tunneling diagrams.

\end{document}